\documentclass[preprint,11pt]{elsarticle}
\usepackage{graphicx,amssymb,amsmath}
\usepackage{epstopdf}
\usepackage[parfill]{parskip}    
\usepackage{url,hyperref}

\usepackage{bm}
\usepackage[numbers]{natbib}
\usepackage{graphics,graphicx,subfig,color}
\usepackage{epstopdf}
\usepackage[parfill]{parskip}    

\newcommand{\eps}{{\varepsilon}}

\newcommand{\ve}{{\varepsilon_0}}

\renewcommand*{\Xi}{{\boldsymbol{\xi}}}

\newcommand*{\E}{{\mathbb{E}}}
\newcommand*{\V}{{\mathbb{V}}}
\newcommand*{\R}{{\mathbb{R}}}

\DeclareMathOperator{\dive}{div}
\newcommand{\pav}[2]{\left\langle #1 \right\rangle_{\varepsilon_#2}}
\newcommand{\ueps}[1]{_{\varepsilon_{#1}}}
\newcommand{\bxi}{{\xi}}
\newcommand{\q}{\mathfrak{q}}
\DeclareMathOperator{\Lip}{Lip}

\allowdisplaybreaks
\journal{Physica D}

\begin{document}
\begin{frontmatter}

\title{On spurious detection of linear response and misuse of the fluctuation-dissipation theorem in finite time series}

\author[USyd]{Georg A. Gottwald}
\ead{georg.gottwald@sydney.edu.au}

\author[USyd]{Caroline L. Wormell}
\ead{ca.wormell@gmail.com}

\author[USyd,MPI]{Jeroen Wouters}
\ead{jeroen.wouters@uni-hamburg.de}

\address[USyd]
{School of Mathematics and Statistics, University of Sydney, NSW 2006, Australia}

\address[MPI]
{Universit\"at Hamburg Geowissenschaften, Meteorologisches Institut, Bundesstr. 55, 20146 Hamburg, Germany}

%
%


\begin{abstract}  
Using a sensitive statistical test we determine whether or not one can detect the breakdown of linear response given observations of deterministic dynamical systems. A goodness-of-fit statistics is developed for a linear statistical model of the observations, based on results for central limit theorems for deterministic dynamical systems, and used to detect linear response breakdown. We apply the method to discrete maps which do not obey linear response and show that the successful detection of breakdown depends on the length of the time series, the magnitude of the perturbation and on the choice of the observable.\\ 
We find that in order to reliably reject the assumption of linear response for typical observables sufficiently large data sets are needed. 
Even for simple systems such as the logistic map, one needs of the order of $10^6$ observations to reliably detect the breakdown with a confidence level of $95\%$; if less observations are available one may be falsely led to conclude that linear response theory is valid. The amount of data required is larger the smaller the applied perturbation. For judiciously chosen observables the necessary amount of data can be drastically reduced, but requires detailed {\em{a priori}} knowledge about the invariant measure which is typically not available for complex dynamical systems.\\ 
Furthermore we explore the use of the fluctuation-dissipation theorem (FDT) in cases with limited data length or coarse-graining of observations. The FDT, if applied naively to a system without linear response, is shown to be very sensitive to the details of the sampling method, resulting in erroneous predictions of the response.
\end{abstract}

\begin{keyword}
linear response theory; fluctuation-dissipation theorem; climate science
\end{keyword}
\end{frontmatter}

\section{Introduction}
\label{sec-intro}
An important question in the study of probabilistic properties of dynamical systems is how to determine the response of a system if subjected to a small perturbation. For example, in climate science we would like to know how the global mean temperature changes upon increasing ${\rm{CO}}_2$ levels. This problem was solved in statistical physics in the context of thermostatted Hamiltonian systems, establishing the framework of linear response theory \cite{Kubo66,Balescu1,Zwanzig,MarconiEtAl08}. In essence, linear response theory employs a Taylor expansion of the perturbed invariant measure around the unperturbed equilibrium measure; this then allows to calculate averages of observables in the perturbed system entirely from knowledge of the statistics of the unperturbed system.\\
The study of linear response involves two issues: proving differentiability of the response and finding an expression for the derivative of the response. To establish linear response, the invariant measure needs to be differentiable with respect to the parameter describing the magnitude of the perturbation. For the existence of an analytical formula for the response in terms of the equilibrium fluctuations of the unperturbed system, which is the statement of the celebrated fluctuation-dissipation theorem (FDT), the invariant measure needs additionally to be differentiable with respect to the phase space variables.\\
Applying this framework to deterministic dynamical systems, in particular to forced dissipative systems whose dynamics evolves on an attractor of zero Lebesgue measure in the full space, has been a challenge. In a series of papers, Ruelle showed that the response is linear for the class of uniformly hyperbolic Axiom A systems, i.e. the invariant measure is differentiable with respect to the magnitude of the perturbation \cite{Ruelle97,Ruelle98,Ruelle09a,Ruelle09b}.

Due to the singular nature of the invariant measure of forced dissipative systems the fluctuation-dissipation theorem, however, cannot hold. Heuristically this failure can be understood by realizing that typical perturbations will have a non-zero projection along the stable manifold, generally transverse to the attractor, whereas the invariant measure is supported entirely on the attractor. Therefore one cannot estimate the response by solely considering correlations of the unperturbed system. A linear response formula can still be expressed, but involves the full linear tangent dynamics and must take into account the evolution of exponentially attenuated perturbations along stable directions rather than just the unperturbed fluctuations along the unstable manifolds as in the FDT.\\
The hope that linear response theory can be extended to more general chaotic dynamical systems has been dampened by numerical results on the tent map \cite{Ershov93} and rigorous analysis by Baladi and co-workers \cite{BaladiSmania08,BaladiSmania10,Baladi14,BaladiEtAl15,DeLimaSmania15}. In particular, it was shown that the logistic map does not obey linear response. This is due to the non-smooth changes of the invariant measure when perturbing from a chaotic parameter value to a periodic one or vice versa. Even worse, even when restricting to the Cantor set of chaotic parameter values the measure is not differentiable in the sense of Whitney. On the other hand, there are numerical simulations suggesting that linear response might exist for some examples of non-uniformly hyperbolic systems \cite{Reick02,CessacSepulchre07,LucariniSarno11} including the Lorenz '63 system which involves homoclinic tangencies. Furthermore, the lack of structural stability, which was believed to be an obstruction to linear response theory in the climate system \cite{McWilliams07}, does not preclude the existence of linear response as was rigorously shown in \cite{Dolgopyat04}. 
The current belief in the mathematical community is that a sufficient condition for the existence of linear response is the summability of the correlation function; the summability of the correlation function is, however, shown not to be necessary for general observables  \cite{Korepanov15,BaladiTodd15}.\\

Notwithstanding the lack of rigorous mathematical proofs for its validity for general forced dissipative non-equilibrium systems, linear response theory has been taken up in the climate sciences to predict the response of the climate, as was first proposed by Leith \cite{Leith75}. 
%
%
Linear response theory and the fluctuation-dissipation theorem have since been used with some success by several groups. They have been applied to various toy models related to atmospheric chaos \cite{MajdaEtAl10,LucariniSarno11,AbramovMajda07,AbramovMajda08,CooperHaynes11,CooperEtAl13}, barotropic models \cite{Bell80,GritsunDymnikov99,AbramovMajda09}, quasi-geostrophic models \cite{DymnikovGritsun01}, atmospheric models \cite{NorthEtAl93,CionniEtAl04,GritsunEtAl02,GritsunBranstator07,GritsunEtAl08,RingPlumb08,Gritsun10} and coupled climate models \cite{LangenAlexeev05,KirkDavidoff09,FuchsEtAl14,RagoneEtAl15}.
%
These successes have led scientists to believe that high-dimensional complex systems may very well obey linear response. The standard argument is that complex systems involve a multitude of interacting processes on several temporal and spatial scales and behave effectively stochastically with a smooth invariant measure \cite{MajdaEtAl10}. This point of view seems at least reasonable for observables of the slow dynamics of complex multi-scale systems which in the limit of infinite time-scale separation are asymptotically stochastic \cite{MelbourneStuart11,GottwaldMelbourne13c,KellyMelbourne14}. In the case of stochastic dynamical systems linear response theory can indeed be justified \cite{Haenggi78,HairerMajda10}. However, several instances are now known where atmospheric and oceanic dynamics exhibits a rough dependence on parameters \cite{ChekrounEtAl14}, and where, even if linear response theory is observed, the fluctuation-dissipation theorem is not valid \cite{CooperHaynes13}.\\
%

On a more fundamental level, however, it is by no means clear that high-dimensional complex systems do obey linear response theory. In this paper we do not attempt to answer this question. Rather, we consider the following practical issue: systems which do not obey linear response theory are observed with {\em{finite}} time series. In such cases we seek to show that the breakdown might not be detectable, and the system's observed behavior may appear consistent with linear response theory. Moreover, the choice of the observable is crucial for the detectability of the breakdown of linear response in finite time series. In particular, we will show that global observables are less able to detect the non-smoothness of the invariant measure whereas local observables which hone in on the roughness of the invariant measure  will make the non-smoothness apparent for smaller amounts of data. Finally, the perturbation size also impacts on the detectability of breakdown, with smaller perturbations requiring more data for successful breakdown detection. \\

This work is motivated by the contradiction between the reported success of linear response theory in the climate sciences and rigorous mathematical results proving the non-existence of linear response theory for a large class of dynamical systems.\\

The paper is organized as follows. In Section~\ref{sec-lrt} we briefly review linear response theory and the fluctuation-dissipation theorem. In Section~\ref{sec-eq} we propose a goodness-of-fit test to probe for the validity of linear response in time series. In Section~\ref{sec-break} we discuss the logistic map, demonstrate the mechanism leading to the breakdown of linear response for this one-dimensional map and show how this breakdown might not be apparent with time series of insufficient length. We show the effect of finite data size as well as how the choice of the observable can either mask or emphasize the non-smoothness of the invariant measure. In Section~\ref{sec-fdt} we show further that an application of the FDT in situations where linear response does not exist cannot provide any reliable statistical information, not even in an averaged sense. We conclude with a summary in Section~\ref{sec-summary}.

\section{Linear response theory}
\label{sec-lrt}
We consider here a family of dynamical systems $f_\varepsilon:M \to M$ on some space $M$. We assume that the map $f_\varepsilon$ depends smoothly on the parameter $\varepsilon$ and that for each $\varepsilon$ the dynamical system admits a unique invariant physical measure $\mu_\varepsilon$, e.g. absolutely continuous measures or Sinai-Ruelle-Bowen measures (SRB). An ergodic measure is called physical if for a set of initial conditions of nonzero Lebesgue measure the temporal average of a typical observable converges to the spatial average over this measure. Considering an observable $A:M\to\R$, we are interested in the change of the average of the observable 
\begin{align*}
\langle A\rangle_\varepsilon = \int_M A\, d\mu_\varepsilon
\end{align*}
upon varying $\varepsilon$. A system is said to have {\em{linear response}} if the derivative
\begin{align*}
\langle A\rangle^\prime_{\varepsilon_0}:= \frac{\partial}{\partial\varepsilon} \langle A\rangle_\varepsilon |_{\varepsilon_0}
\end{align*}
exists. It is obvious that a sufficient condition for linear response is that the invariant measure $\mu_\varepsilon$ is differentiable with respect to $\varepsilon$. If the limit does not exist, we say there is a breakdown of linear response. We assume that the observable captures sufficient dynamic information about the dynamical system; for example, an odd observable on a system symmetric about $0$ would be identically zero regardless of whether the system had a linear response or not.\\

One may further ask whether, if linear response exists, a computable analytical expression for the linear response 
\begin{align}
\langle A\rangle_\varepsilon \approx \langle A\rangle_{\varepsilon_0} + \langle A\rangle^\prime_{\varepsilon_0}\,\delta \varepsilon
\label{e.linresp}
\end{align}
can be found for small values of $\delta \varepsilon=\varepsilon-\varepsilon_0$. To write down an expression of the linear response, we introduce a vector field $X$ as $X\circ f_{\varepsilon_0} :=\partial_{\varepsilon} f_\varepsilon |_{\varepsilon=\varepsilon_0}$. Note that the introduction of the vector field $X$ is the standard way of formulating perturbations in statistical physics as $f_\varepsilon = f_{\varepsilon_0}+\delta\varepsilon X(f_{\varepsilon_0})$. The linear response $\langle A\rangle^\prime_{\varepsilon_0}$ can then be formally expressed as
\begin{align}
\label{e.fdt0}
\frac{\partial}{\partial\varepsilon} \langle A\rangle_\varepsilon |_{\varepsilon_0}
= \sum_{n = 0}^{\infty} \langle X (x) \nabla (A \circ
  f_{\varepsilon_0}^n) (x)\rangle_{\varepsilon_0} ,
\end{align}
for $x\in M$. Provided the unperturbed invariant measure $\mu_{\varepsilon_0}$ has a density $\rho_{\varepsilon_0} (x)$ that is differentiable with respect to $x\in M$ and non-vanishing, one can perform partial integration in (\ref{e.fdt0}) to rewrite the linear response in terms of an integral of a correlation function. This form of the linear response formula is known as the fluctuation-dissipation theorem \cite{Ruelle98,MarconiEtAl08} and reads as
\begin{align}
\label{e.fdt}
\frac{\partial}{\partial\varepsilon} \langle A\rangle_\varepsilon |_{\varepsilon_0}
= -\sum_{n = 0}^{\infty}
  \left\langle \left( \frac{\nabla (\rho_{\varepsilon_0} (x) X
  (x))}{\rho_{\varepsilon_0} (x)} \right) A \circ f^n_{\varepsilon_0} (x)
  \right\rangle_{\varepsilon_0}\; .
\end{align}
In the form (\ref{e.fdt}) the response formula is easier to apply to a numerical integration or to experimental data than the original response formula (\ref{e.fdt0}), as it can be estimated directly from a long integration. The assumption that the invariant measure is differentiable is, however, a strong limitation, as it fails for dissipative systems with singular measures with support on an attractor as well as for absolutely continuous maps involving singularities such as the logistic map (see Section~\ref{sec-break}).\\

Introducing the notation of a divergence operator with respect to a density $\rho$
\begin{align*} 
\dive_\rho{B}\,(x) = \frac{\dive{(\rho B)}(x)}{\rho(x)}, 
\label{e.divrho} 
\end{align*}
the response formula (\ref{e.fdt}) can be concisely written as
\begin{align}
\frac{\partial}{\partial\varepsilon} \langle A\rangle_\varepsilon|_{\varepsilon_0} = - \sum_{n=0}^{\infty} C_n(\dive_{\rho_{\varepsilon_0}}{X},A)\, ,
\end{align}
with the correlation function $C_n$ between two observables $A$ and $B$ defined as
\begin{align*}
C_n(A,B) = \pav{{A \hphantom{;}B\circ f^n}}{0} -\, \pav{A}{0} \pav{B}{0} .
\end{align*}
For sufficiently fast decay of correlations one can estimate (\ref{e.fdt}) from a time series $x_{i = 1,\dotsc,N}$ of finite length $N$ via
\begin{align}
\frac{\partial}{\partial\varepsilon} \langle A\rangle_\varepsilon |_{\varepsilon_0}  \approx -\sum_{n=0}^{n_{\rm{max}}} \frac{1}{N-n} \sum_{i=1}^{N-n} \left(\dive_{\rho_{\varepsilon_0}}{X}\right)(x_i) A(x_{i+n}) 
\label{e.finitedatafdt} 
\end{align}
with $1\ll n_{\rm{max}}\ll N$. This expression allows for the estimation of the first-order response to a perturbation using a times series of the unperturbed system, provided the unperturbed density $\rho_{\varepsilon_0}$ can be estimated from the time series as well. In the climate sciences $\rho_{\varepsilon_0}$ is mostly approximated either via a quasi-Gaussian approximation \cite{GritsunBranstator07,GritsunEtAl08} or by kernel smoothing \cite{CooperHaynes11}.\\

\begin{figure}[htbp]
\centering
\includegraphics[width=0.7\linewidth]{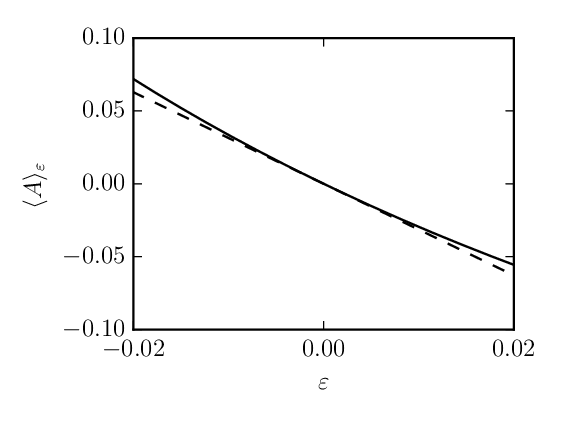}
\caption{Response of the doubling map to the perturbation $X(x) = \sin 2\pi x$ for an observable $A(x) = \cos 2\pi x$. The continuous line represents the actual response, the dashed line depicts the result of the fluctuation-dissipation theorem (\ref{e.finitedatafdt}). Note that for small values of the perturbation $\varepsilon$ the two curves are indistinguishable.}
\label{fig:spdresp}
\end{figure}

Before exploring examples where linear response does not exist, we show in Figure~\ref{fig:spdresp} an example of linear response for the doubling map $f_\varepsilon(x)={\rm{mod}}(2 x + \varepsilon \sin(4 \pi x),1)$ which for $\varepsilon=0$ admits the Lebesgue measure as its invariant measure. Here the fluctuation-dissipation formula (\ref{e.finitedatafdt}) becomes
\begin{align*}
\frac{\partial}{\partial\varepsilon}\langle A\rangle_\varepsilon |_{\varepsilon_0}= -\pi
\end{align*}
and accurately reproduces the actual response. We estimate the actual response numerically using a spectral method. In particular, we approximate the transfer operator which propagates densities under the perturbed dynamics $f_\varepsilon$ (see, for example, \cite{Baladi}) by projecting onto a finite number of basis function\cite{Ding93,Boyd01,Trefethen13}. For the doubling map we choose $100$ trigonometric functions. The invariant measure $\rho_\varepsilon$ is then approximated by the eigenfunction corresponding to the eigenvalue $1$ of the approximated transfer operator. The advantage of spectral methods over using a long but finite time series with subsequent binning is its high accuracy and fast convergence with the number of resolved eigenfunctions \cite{Trefethen13}. Their applicability, however, is restricted to low-dimensional systems. 

\section{Testing for linear response in finite time series}
\label{sec-eq}
In this section we develop a quantitative goodness-of-fit test for the detectability of linear response in time series of finite size which allows to make statements about the significance of an observed linear response.
Given a family of chaotic maps $f_\varepsilon$ that may or may not obey linear response, we test for linear response at some reference state with parameter $\varepsilon=\varepsilon_0$ by examining the linear dependency of the response
\begin{align}
\delta A = \langle A\rangle_\varepsilon - \langle A\rangle_{\varepsilon_0}
\end{align}
for $M > 2$ different values of the perturbation parameter $\varepsilon_1,\dotsc,\varepsilon_M$, by sampling $N_1,\dotsc,N_M$ consecutive values from the equilibrium dynamics of $f\ueps{1},\dotsc,f\ueps{M}$, respectively. Explicitly, for each $i=1,\dotsc, M$, we have time series $x^i_n=f\ueps{i}(x^i_{n-1})$ for $n= 1, \dotsc, N_M$. The initial conditions $x^i_0$ are distributed according to the physical measure associated with $f\ueps{i}$.\\ We consider bounded and continuous observables and assume that for each member of the family $f_{\varepsilon_i}$ the autocorrelation $C_j(A,A)$ decays sufficiently rapidly, and that the lengths of the time series $N_i$ are large compared to typical decay times of the autocorrelation function; in practice we choose $N_i \gg \tau_{A,\varepsilon_i}$, where $\tau_{A,\varepsilon_i}$ is the $1/e$-folding time of $A$ under the dynamics $f_{\varepsilon_i}$. We further set, for simplicity, $N_i=N$ for all $i$.\\

For a large class of chaotic dynamical systems, the sample averages of the observations
\begin{align}
\bar{A}_i=\frac{1}{N_i}\sum_{n=1}^{N_i}A(x^i_n)
\label{e.Aavgs}
\end{align}
obey the central limit theorem and are distributed asymptotically as $\mathcal{N}\left(\pav{A}{{i}}, \sigma_i^2/N_i\right)$ \cite{melbournelectures,ColletEckmann07}.
The variances $\sigma_i^2$ are given by the Green-Kubo formula in terms of lag-correlations of $f\ueps{i}$ as
\begin{align} 
\sigma_i^2 = C_0(A,A) + 2 \sum_{j=1}^{\infty} C_j(A,A)\; . 
\label{e.covar}
\end{align}
Numerically, the variances are determined as a Monte-Carlo estimate from observations of the observables under the perturbed dynamics using the central limit theorem. According to the central limit theorem
\begin{align}
\bar{A}_i = \langle A\rangle_{\varepsilon_i}  + \frac{\sigma_i}{\sqrt{N}} \xi_i \; ,
\label{e.clt}
\end{align}
for $i = 1,\dotsc,M$ and {\em{iid}} noise $\xi_i \sim N(0,I)$. If the dynamical system indeed has linear response and provided the perturbations $\delta \varepsilon_i=\varepsilon_i-\varepsilon_0$ are sufficiently small, the following statistical model holds for $\bar{A}_i$
\begin{align}
\bar{A}_i  = \alpha_0 + \alpha_1 \, \delta\varepsilon_i + \frac{\sigma_i}{\sqrt{N}} \xi_i \, ,
\label{e.linmod}
\end{align}
with $\alpha_0=\langle A\rangle_{\varepsilon_0}$ and $\alpha_1= \langle A\rangle_{\varepsilon_0}^\prime$ for some unperturbed reference state with $\varepsilon=\varepsilon_0$. Note that the $\xi_i$ are independent as the samples from each perturbed system are independent. 

To determine the parameters $\alpha_0$ and $\alpha_1$ of the model (\ref{e.linmod}) from time series we apply a weighted least squares fit to obtain
\begin{align*}
  \left(\begin{array}{c}
    \hat \alpha_0\\
    \hat \alpha_1
  \end{array}\right) = (D^T D)^{- 1} D^T Y
\end{align*}
with the design matrix
\begin{align*}
  D = \left(\begin{array}{cc}
    1 / \sigma_1 & \delta \varepsilon_1 / \sigma_1\\
    \vdots & \vdots\\
    1 / \sigma_M & \delta \varepsilon_M / \sigma_M
  \end{array}\right)\, ,
\end{align*}
and the vector of scaled observations
\begin{align*}
  Y = \left(\begin{array}{c}
   \bar{A}_1 / \sigma_1\\
    \vdots\\
   \bar{A}_M / \sigma_M
  \end{array}\right)\, .
\end{align*}
Higher-order responses can naturally be incorporated by adding a quadratic term $\alpha_2\,\delta\varepsilon_i^2$ to (\ref{e.linmod}) and employing higher-order regression allowing, in principle, for a larger range of perturbations (in case linear response exists).\\

To test whether the observations could have been drawn from the linear model (\ref{e.linmod}) with normally distributed errors $\xi_i$ with mean zero and variance $1$, we choose a Pearson $\chi^2$-test to test the goodness-of-fit with statistics
\begin{align}
\chi^2 
&= N\, \sum_{i=1}^M \left(Y_i - \frac{1}{\sigma_i}\left(\hat \alpha_0+\hat \alpha_1\varepsilon_i\right)\right)^2 
\nonumber
\\
&= N\, Y^T(I-H)Y,
\label{e.chisq}
\end{align}
where the idempotent hat matrix 
\begin{align*}
H = D (D^T D)^{-1} D^T
\end{align*} 
maps scaled observations $Y$ to their linear fits, i.e. $HY = D(\hat \alpha_0\;\;\hat \alpha_1)^T$ \cite{BoxHunterHunter}.\\
If the response of the underlying dynamical system is linear, $\chi^2$ has a $\chi^2$-distribution with $M-2$ degrees of freedom and expectation value $\E \chi^2_{M-2} = M-2$. 
We therefore introduce as a measure for the breakdown of linear response the difference between the $\chi^2$ test statistic for the scaled observations $Y_i= \bar{A}_i / \sigma_i$ and the expectation of the test statistic under the null hypothesis of linear response
\begin{align} 
\q= \frac{1}{N}\left( \chi^2 -\E \chi^2_{M-2} \right).
\label{e.Echi2_0}
\end{align}
Defining $W$ as the vector with components $W_i=\langle A\rangle_{\varepsilon_i}/\sigma_i$ we can use the central limit theorem (\ref{e.clt}), which holds independent of the existence of linear response, to obtain the following expressions for the mean and variance of the breakdown parameter. The mean is calculated as
\begin{align} 
\E \q&= \frac{1}{N}\left(\E\chi^2 -\E \chi^2_{M-2}\right)
\nonumber
\\ 
&=  \E\bigg((W + \frac{1}{\sqrt{N}}\bxi)^T(I-H)(W + \frac{1}{\sqrt{N}}\bxi) - \frac{1}{N}\E \chi^2_{M-2}\bigg)
\nonumber
\\
&= \|W-HW\|^2,
\label{e.Echi2}
\end{align}
where we used that $H$ is idempotent. Hence $\q$ is a random variable whose expected value measures the difference between the actual response $\langle A\rangle_{\varepsilon_i}$ and an assumed linear response $\alpha_0+\alpha_1\varepsilon_i$ as calculated via least square regression. We have $\E \q\ge 0$ with equality only for $W = HW$, i.e. if the actual response is linear. The variance of the breakdown parameter $\q$ is calculated as
\begin{align*} 
\V \q &= \E \bigg((W + \frac{1}{\sqrt{N}}\bxi)^T(I-H)(W + \frac{1}{\sqrt{N}}\bxi) -\frac{M-2}{N} - \E \q \bigg)^2\\
&= \frac{1}{N} \E\left( \bxi^T (I-H) (2W + \frac{1}{\sqrt{N}}\bxi)  - \frac{M-2}{\sqrt{N}}  \right)^2.
\end{align*}
This shows that $\q$ is a consistent estimator for the mismatch $\E\q = \|W-HW\|^2$ since $\V\q\to 0$ for $N\to \infty$. In the numerical experiments in Section~\ref{sec-finite} we will consider Monte-Carlo estimates of the mismatch over realizations $\q_j$ differing in their initial condition and set
\begin{align}
\hat\q=\frac{1}{K} \sum_{j=1}^K\q_j\, .
\end{align}

Now, consider a dynamical system which does not obey linear response, i.e. $\E\q\neq 0$. Using Chebyshev's inequality we have that for all $b < N\E\q$,
\begin{align*}
P(N\q < b) &\leq P(|\q-\E\q| > \E\q - b/N)\\
&\leq \frac{\V(\q)}{(\E\q - b/N)^2}\; .
\end{align*}
Since, as we have shown above, $\V \q\to 0$ as $N\to\infty$, we conclude that $N \q \to \infty$ in probability as $N\to\infty$. Hence, if $F$ is the cumulative distribution function of the $\chi^2_{M-2}$ distribution, the $p$-value obtained using the $\chi^2$- test
\begin{align}
p  = 1 - F(\chi^2) = 1- F(M-2 +N\q)
\label{e.p}
\end{align}
converges quickly in probability to zero as $N\to\infty$ \cite{BoxHunterHunter}. In practical terms this means that the probability of falsely accepting the null hypothesis of linear response at any significance level can be made arbitrarily small if $N$ is large enough.


For a specified significance level $\alpha$ we can now define
\begin{align}
\label{e.qalpha}
\q_\alpha = \frac{1}{N} \left(F^{-1}(1-\alpha)-(M-2)\right)\, .
\end{align}
This defines a threshold value for the observed random variable $\hat\q$ such that if $\hat \q >\q_\alpha$ the null hypothesis of linear response is rejected with significance level $\alpha$ (i.e. with probability $1-\alpha$); conversely, if $\hat \q <\q_\alpha$ the null hypothesis of linear response is accepted with significance level $\alpha$ (i.e. with probability $1-\alpha$).

The detectability of breakdown of linear response is linked to the amount of available data. As $N\to \infty$, a breakdown will always become detectable at any specified significance level $\alpha$. Conversely, if the mismatch $\E\q$ between the true response of the dynamical system and the linear response is too small and there is an insufficient amount of data available, the actual response will be swamped by the sampling noise, and one will not be able to detect the breakdown of linear response with a reasonable significance level.

In Section~\ref{sec-finite} we will use our goodness-of-fit test to study the detectability of breakdown of linear response in time series of finite length.




\section{Breakdown of linear response theory}
\label{sec-break}
A standard dynamical system for which linear response fails \cite{Baladi14} is the logistic map $f:[0,1]\to[0,1]$ given by
\begin{align}
f(x)=ax(1-x)\,
\label{e.log}
\end{align}
for $a\in[0,4]$. This family of maps is particularly well-understood~\cite{Lyubich02,AvilaMoreira05}: we can decompose the parameter interval according to $[0,4]=\mathcal{P}\cup\mathcal{C}\cup\mathcal{N}$ where $\mathcal{N}$ has Lebesgue measure zero, and the asymptotic dynamics consists of a periodic attractor for $a$ in the open and dense set $\mathcal{P}$ and of a strongly chaotic attractor for $a$ in the set $\mathcal{C}$ of positive measure. For $a\in \mathcal{C}$ the logistic map admits a unique absolutely continuous invariant measure (a.c.i.m.) \cite{Jakobson81,Benedicks85} and moreover satisfies the Collet-Eckmann condition~\cite{ColletEckmann83}, i.e. there exists $d>0$, $\lambda>1$ such that 
\[
|Df^n(f({\textstyle\frac12}))| \ge d\lambda^n\enspace\text{for all $n\ge1$}.
\]
Then the Lyapunov exponent is positive, the attractor $\Lambda$ consists of finitely many intervals $\Lambda_1,\dots,\Lambda_q$ permuted cyclically by $f$, and $f^q|_{\Lambda_i}$ has exponential decay of correlations for H\"older observables for each $i=1,\dots,q$~\cite{KellerNowicki92,Young02}.\\

\begin{figure}[htbp]
\centering
\includegraphics[width=0.7\linewidth]{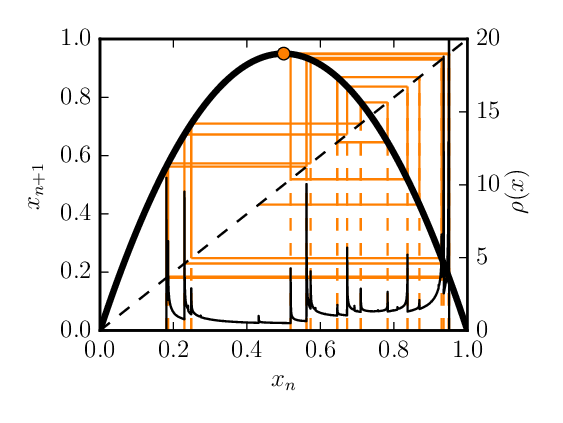}
\caption{The a.c.i.m. of the logistic map (\ref{e.log}) with $a = 3.8$ (black lines), and a cobweb diagram (online orange) relating the spikes of the a.c.i.m. to the first $17$ iterates of the forward orbit of the critical point.}
\label{fig.l-copm}
\end{figure}
\begin{figure}[htbp]
\centering
\includegraphics[width=0.9\linewidth]{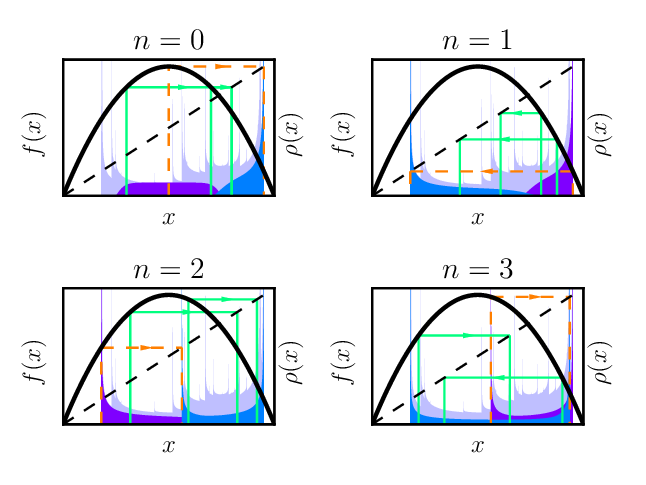}
\caption{The repeated action of the logistic map (\ref{e.log}) with $a = 3.8$ on a continuous density (blue), with a.c.i.m. (light blue). The  dashed lines (online orange) show the image of the critical point after $n$ iterations of the logistic map. The continuous lines (online green) show the image of a few points within the support of the initial density at $n=0$ upon subsequent iteration of the map.}
\label{fig.logprop}
\end{figure}

The logistic map is not uniformly expanding and has a critical point at $x=c=1/2$ with $f'(c) = 0$. The critical point gives rise to a complicated and rough absolutely continuous invariant measure, because $f$ and its iterates $f^n$ compress the phase space around $x=1/2$ non-uniformly (see Figure~\ref{fig.l-copm}). We summarize here the analysis given in \cite{Ruelle09b}. Near the critical point $c$ we approximate $y=f(x) \approx c_1+\tfrac12 f^{\prime\prime}(c)(x-c)^2$  with $c_n=f^n c$ and hence $x-c=\pm \sqrt{2(c_1-y)/f^{\prime\prime}(c)}+O(b-y)$. This implies that an initial smooth density $\rho_0(x)$ including the critical point $x=c$ in its support will evolve under the dynamics into a spike with a square-root singularity at $x=c_1$. Propagating the density for a further time step will transport this peak to $x=c_2$ and create a second, new spike at $x=c_1$, and so forth. The expanding action of the logistic map away from the critical point leads to a broadening of the spikes, and thereby consecutive spikes will have smaller amplitudes, preserving the normalization of the initial density $\rho_0(x)$. This is illustrated in Figure~\ref{fig.logprop} and can be formalized to find an explicit formula for the unique a.c.i.m. in terms of its density 
\begin{align}
\rho(x) = \phi(x) + \sum_{n=1}^{\infty} \eta_n(x). 
\label{e.loga.c.i.m.}
\end{align}
Here $\phi(x)$ represents a continuous background density with $\phi(c_1)=\phi(c_2)=0$. The countably infinite family of spikes  $\eta_n$ are found to be
\begin{align}
\eta_n(x) \sim \Upsilon_n \frac{1}{\sqrt{x-f^n c}},
\label{e.spike}
\end{align}
with magnitude 
\begin{align}
 \Upsilon_n =  \rho(c)\left|\frac{1}{2} f^{\prime\prime}(c)\prod_{i=1}^n f^\prime(c_i)\right|^{-\frac{1}{2}}.
\label{e.spikem}
\end{align}
For large $n$ the product in \eqref{e.spikem} is asymptotically $\alpha^n$, where $1<\alpha<2$ denotes the Lyapunov multiplier, hence the magnitude $\Upsilon_n$ of the spikes decays as $\alpha^{-n/2}$. This implies that the widths of the spikes (defined as the distance from the singularity at $x=f^nc$ at which the amplitudes drop to some chosen threshold) scale like $\alpha^{-n}$.

\begin{figure}[htbp]
\centering
\includegraphics[width=0.7\linewidth]{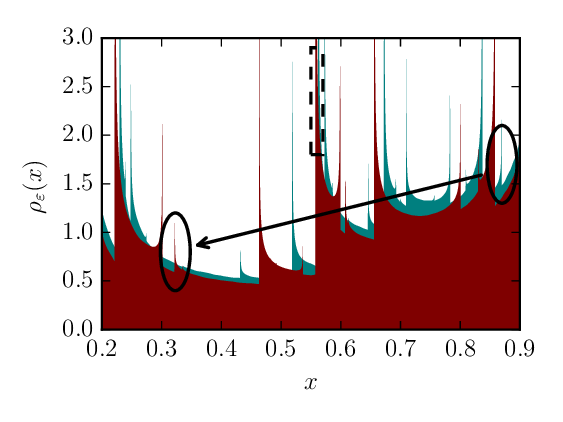}
\caption{Overlaying two a.c.i.m.s of the logistic map with $a=3.8$ (maroon) and $a=3.8+6.05\times 10^{-4}$ (light blue). The displacement of a smaller spike with $n=17$ is illustrated by the two ovals, and that of a large spike with $n=3$ with a dashed box.}
\label{fig.lcpet}
\end{figure}

\begin{figure}[htbp]
\centering
\includegraphics[width=0.7\linewidth]{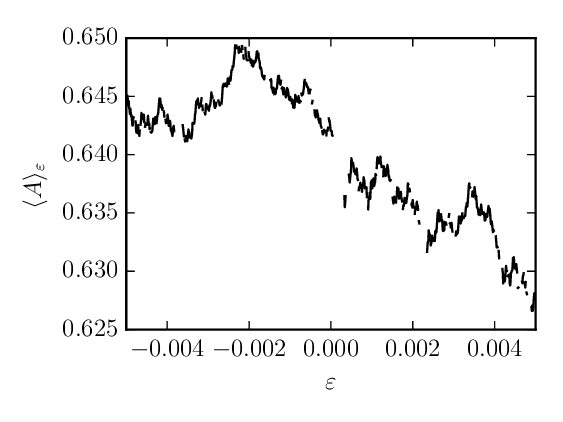}
\caption{Observable $\langle A\rangle_\eps$ as a function of the perturbation $\eps$ for the logistic map (\ref{e.log}) with $a=3.8(1+\varepsilon)$ and observable $A(x) = x$. Gaps in the curve correspond to periodic windows.}
\label{fig.logresp}
\end{figure}

We will now study the effect of parameter perturbations $a=a_0(1+\varepsilon)$ onto the logistic map (\ref{e.log}) and its a.c.i.m. $\rho(x)dx$.  We may ask how fast the spikes move upon increasing $\eps$. Expanding the displacement length $\ell_\eps = \left| f_\varepsilon^{n+1}(c) - f_0^{n+1}(c) \right|$ around $\eps=0$ yields that the speed $v_n = \ell_\eps / \eps$ is proportional to $\partial_\varepsilon f_\varepsilon^{n+1}(c)= \prod_{i=1}^nf^\prime(c_i) + O(\varepsilon)$ and hence is proportional  to $\alpha^n$ for large $n$. Hence the smaller spikes move faster than the larger spikes corresponding to small values of $n$. This is illustrated in Figure~\ref{fig.lcpet} where we overlay the invariant densities corresponding to a small perturbation with $\varepsilon= 6.05\times 10^{-4}$. The family of perturbed invariant measures can then be formally written in terms of their associated densities as
\begin{align*}
\rho_{\eps}(x) = \phi_\eps(x) + \sum_{n=1}^{\infty} \eta^{(\eps)}_n(x + \varepsilon \alpha^n), 
\end{align*}
where the spikes are given as in (\ref{e.spike})-(\ref{e.spikem}) with $f$ replaced by $f_\eps$, and the magnitude of the perturbed spikes also decays as $\Upsilon_n^{(\eps)}\sim\alpha^{-n/2}$. Differentiation of $\rho_\eps(x)$ with respect to $\eps$ produces an exponentially growing term $\alpha^{n/2}$ inside the sum which prevents the differentiability of the a.c.i.m., and hence causes the breakdown of linear response. A different way to see the non-differentiability of the invariant measure is to consider the linear response of an indicator function with support $[0,c_n]$. Without loss of generality we assume that the spike has support to the right of $c_n$ and moves to the left upon perturbation (if this is not the case, take $\eps \rightarrow -\eps$ to change the direction). Upon applying a perturbation  $\varepsilon$ the spike will enter the support of the observable and the probability mass moving into the interval is proportional to 
\begin{align*}
\delta \rho_{\rm{spike}} \sim \int_0^{\alpha^n \eps} \frac{\alpha^{-\frac{n}{2}}}{\sqrt{x}}dx\sim \sqrt{\eps} ,
\end{align*}
and therefore $\delta\rho_{\rm{spike}}/\eps\sim1/\sqrt{\eps}$. Since spikes are dense on the support of the a.c.i.m., the non-smoothness extends to the whole a.c.i.m. The non-differentiability of the a.c.i.m. is clearly seen in Figure~\ref{fig.logresp} where we show the observable $\langle A\rangle_\eps$ as a function of $\eps$. The results shown are obtained here again employing spectral methods \cite{Ding93,Boyd01,Trefethen13}.
%
%


\section{Resolving breakdown of linear response in finite time series}
\label{sec-finite}
The rigorous theory by Baladi and co-workers \cite{BaladiSmania08,BaladiSmania10,Baladi14,BaladiEtAl15} shows that certain dynamical systems such as the logistic map do not obey a linear response. In this section we will investigate how the finitude of data may prevent the breakdown to be detectable and how one may falsely be led to believe that linear response was valid.  

As seen in Section~\ref{sec-break} the non-smoothness of the invariant measure is caused by the rapid displacement of spikes upon perturbation. The smaller and narrower the spike, the faster it moves. This points to an issue of resolution: the faster spikes carry less mass and therefore require a certain amount of data to be reliably resolved; the slower spikes carry more mass but their smaller displacement upon perturbation requires sufficient data to be resolved. This means that a sufficiently large amount of data is needed for the breakdown parameter $\mathfrak{q}$ to accurately estimate the mismatch $\E\q$, and to determine whether a system obeys linear response or not. This issue of resolving the mismatch $\E\q$ is an additional finite size issue to the one discussed in Section~\ref{sec-eq} whereby $N$ needs to be sufficiently large to assure that the observed $p$-value is properly estimated (cf. (\ref{e.p})).


Throughout the paper we simulate the logistic map (\ref{e.log}) with $a=a_0(1+\eps)$ and $a_0= 3.8$. In the notation of Section~\ref{sec-lrt} we set $\varepsilon_0=0$ from now on. We choose $M=20$ equidistant values $\eps_i = \frac{2i-M+1}{2}d\eps$ with $i=0,\ldots,M-1$ for some $d\eps>0$ to determine the breakdown parameter $\q$. Note that the breakdown parameter cannot be determined at an exact perturbation size $\eps$ but we determine the validity of a linear approximation over a range of perturbation sizes parametrized by $d\eps$ (for fixed $M=20$). We restrict the set of perturbations $\eps_i$ to include only those which belong to the chaotic Cantor set $\mathcal{C}$. 

The variances $\sigma_i^2$ (\ref{e.covar}) are estimated as a Monte-Carlo estimate from the observed response using the central limit theorem (\ref{e.clt}) with $200$ realizations for each $i=1,\cdots,M$.  


\subsection{Effect of finite data length $N$}

\begin{figure}[htbp]
\centering
\includegraphics[width=0.7\linewidth]{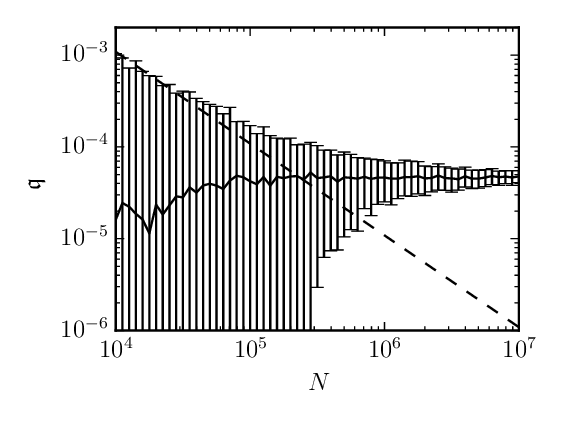}\\
\includegraphics[width=0.7\linewidth]{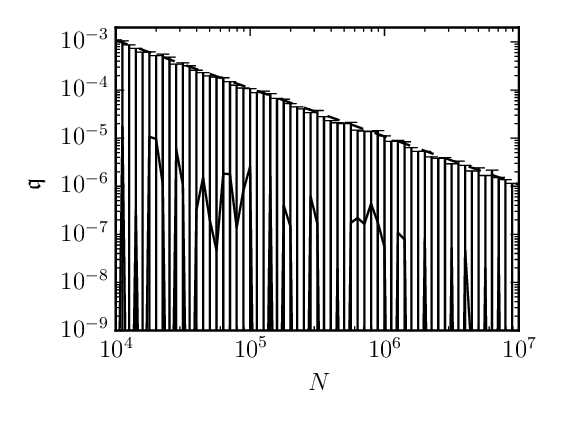}
\caption{Breakdown parameter $\hat{\q}$ (solid line) as a function of the data size $N$, estimated using $d\eps = 10^{-6}$. Top: logistic map (\ref{e.log}) with fixed range of perturbations $a=3.8(1+\varepsilon)$ and observable $A(x) = x$. Bottom: doubling map with perturbation $X(x) = \sin 2\pi x$ and observable $A(x) = \cos 2\pi x$.\\ The error bars show the two-sided $90\%$ prediction interval for $\q$ as estimated from $K=200$ realizations differing in the initial conditions. The dashed line shows $\q_\alpha$ for $\alpha=0.05$. Note that for the doubling map (bottom) the breakdown parameter $\hat{\q}$ assumes values below the plotted range for some values of $N$.}
\label{fig.qN}
\end{figure}

Figure~\ref{fig.qN} shows how the breakdown parameter $\mathfrak{q}$ behaves with increasing data length $N$ for given perturbation size $d\eps=10^{-6}$. For each value of $N$ the breakdown parameter is calculated for the above mentioned range of $M=20$ perturbation sizes $\eps_i$. Shown is the Monte-Carlo estimate $\hat\q$ of the expectation value of the breakdown parameter $\E\mathfrak{q}$ over $K=200$ realizations, differing in the initial condition of the logistic map as well as in the threshold value $\q_\alpha$ corresponding to a significance level $\alpha=0.05$. We show error bars obtained from the ensemble statistics indicating the two-sided $90\%$ prediction interval for $\q$.  
We see clearly the saturation of the breakdown parameter with increasing data length $N$ towards the deterministic limit $\E\q$ which eventually leads to detection of the breakdown above a significance level of $p=0.05$. 
The breakdown can, however, only be detected reliably with a statistical significance level larger than $0.05$ for long time series with $N>600,000$. The corollary of this is that when analyzing single time series of length $N<600,000$ at several values of the perturbation size $\eps$ the error bars lie below $\q_\alpha$ and the dynamics may be falsely classified as obeying linear response.\\ For comparison we have included in Figure~\ref{fig.qN} a plot showing the breakdown parameter as a function of $N$ for the doubling map which does obey linear response with $\E\q \to 0$ for $\eps\to 0$. Here the observed breakdown parameter $\hat\q$ decreases with $N$ according to the law of large numbers and the sample statistics is consistent with the two-sided $90\%$ prediction interval for the whole range of $N$. Since the expectation of the breakdown parameter $\E\q$ approaches zero for vanishing perturbation size, the estimator of $\E\q$ is noisy due to sampling errors, and hence may be small and negative\footnote{In the doubling map we have $\E\q\sim{\mathcal{O}}(\varepsilon^4)$ since deviations of $W_i$ from a linear fit are ${\mathcal{O}}(\varepsilon_i^2)$; cf. (\ref{e.Echi2}).}.

\begin{figure}[htbp]
\centering
\includegraphics[width=0.7\linewidth]{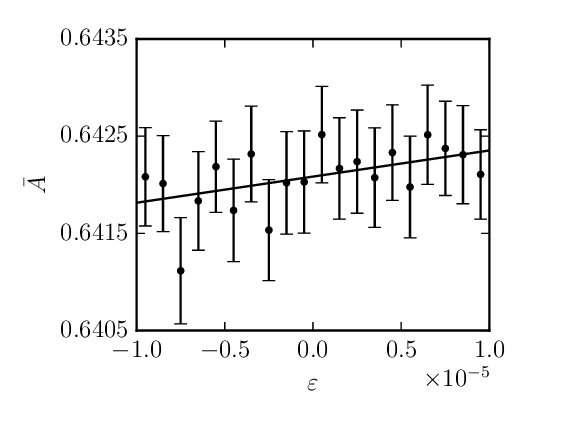}\\
\includegraphics[width=0.7\linewidth]{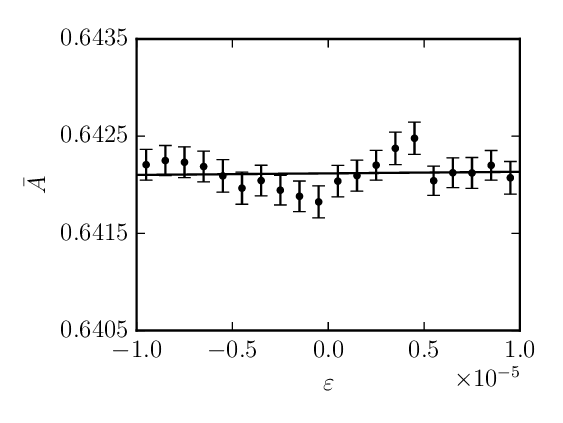}
\caption{The observed sample average $\bar{A}$ as a function of the perturbation size $\eps$ for an observable $A(x) = x$ and a linear fit (solid line). Top: for data length $N=10^5$ where $\hat{\q}=5.03 \times 10^{-5}$ and the breakdown is not detectable ($p=0.148$). Bottom: for data length $N=10^6$ where $\hat{\q}=5.32 \times 10^{-5}$ and the breakdown is detectable ($p=1.31 \times 10^{-8}$). Here $\bar A$ was obtained from a single simulation.}
\label{fig.Aeps}
\end{figure}

A concrete example of how statistical noise may impede the detection of linear response breakdown from observations is shown in Figure~\ref{fig.Aeps} for an observable $A(x)=x$. Shown is the observed sample average $\bar{A}$~\eqref{e.Aavgs} as a function of the perturbation size. The error bars are calculated from the standard deviation as calculated for the single available time series, which is the situation for scientists analyzing observations. For insufficient data length $N=10^5$ a linear response is consistent within the available statistical significance levels (top of  Figure~\ref{fig.Aeps}). Only for significantly larger time series with data length $N=10^6$, does the breakdown become detectable in a statistically significant way (bottom of  Figure~\ref{fig.Aeps}).

%

\subsection{Effect of the perturbation size $\eps$}

\begin{figure}[htbp]
\centering
\includegraphics[width=0.7\linewidth]{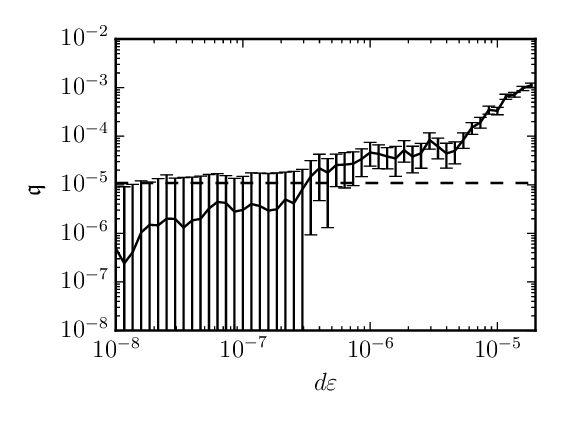}\\
\includegraphics[width=0.7\linewidth]{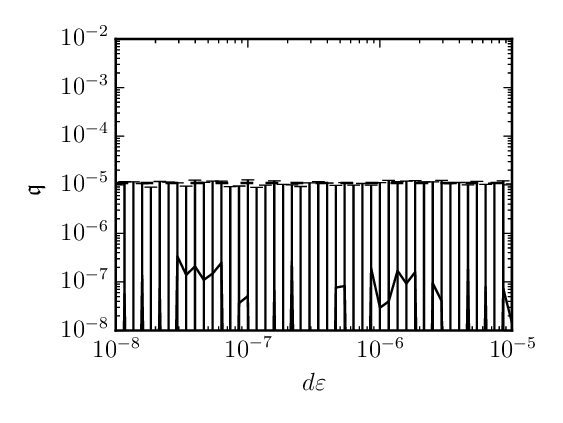}
\caption{Breakdown parameter $\hat{\q}$ (solid line) as a function of the perturbation size $d\eps$ for fixed $N=10^6$.
Top: logistic map (\ref{e.log}) with $a=3.8(1+\varepsilon)$ for an observable $A(x)=x$. Bottom: doubling map with perturbation $X(x) = \sin 2\pi x$ for an observable $A(x) = \cos 2\pi x$.\\ The error bars show the two-sided $90\%$ prediction interval for $\q$ as estimated from $K=200$ realizations differing in the initial conditions. The dashed line shows $\q_\alpha$ for $\alpha=0.05$. Note that for the doubling map (bottom) the breakdown parameter $\hat{\q}$ assumes values below the plotted range for some values of $d\varepsilon$.}
\label{fig.qeps}
\end{figure}

\begin{figure}[htbp]
\centering
\includegraphics[width=0.7\linewidth]{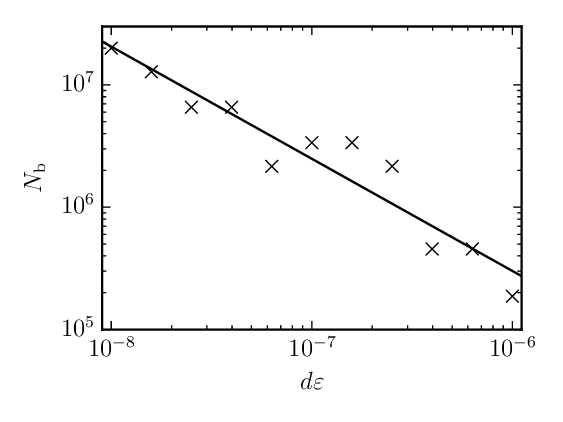}
\caption{The critical data length $N_b$ above which the breakdown of linear response is detectable as a function of the perturbation size $\eps$, parametrized by $d\eps$. The slope of the linear fit is $-0.91$.}
\label{fig.Neps}
\end{figure}

The critical length of the data $N_b$ above which breakdown of linear response can be detected in a statistically significant way depends on the perturbation size $\eps$. 
In particular,  $\chi^2$ is an increasing function of $\eps$ for sufficiently large values of $\eps$, cf. (\ref{e.chisq}).
This dependency can be intuitively understood since the response to small perturbations must be distinguished from the variations in the unperturbed system due to the sampling error. This implies that to be able to identify a deviation from linear response at a specified perturbation size $\eps$ with a significance level $p$ the perturbation size needs to be sufficiently large. This is illustrated in Figure~\ref{fig.qeps} where we show the Monte-Carlo estimate $\hat\q$ of the expectation value of the breakdown parameter as a function of the perturbation size which is parametrized by the perturbation interval $d \eps$. For each value of $d\eps$ the perturbed system is sampled at $\eps_i = \frac{2i-M+1}{2}d\eps$ with $i=0,\ldots,M-1$ for fixed data length $N=10^6$ with $M=20$. For perturbation sizes $d\eps<8\times 10^{-7}$ the observations are consistent with linear response theory and only for $d\eps>8\times 10^{-7}$ can the actual breakdown be detected in a statistically significant way. For comparison we have again included in Figure~\ref{fig.qeps} a plot showing the breakdown parameter as a function of $d\eps$ for the doubling map where linear response assures $\E\q \to 0$ for $\eps\to 0$. Here linear response is consistent with the observations for the whole range of perturbation sizes considered.
Figure~\ref{fig.Neps} illustrates that the smaller the applied perturbation the larger the data length has to be to detect breakdown. Shown is the critical data length $N_b$ above which breakdown can be detected for a given perturbation size. The critical data length $N_b$ was determined to be the value of $N$ such that $\hat{\q} = \q_\alpha$ for $\alpha = 0.05$. A linear fit suggests $N_b \sim \eps^{-\gamma}$, where $\gamma$ was estimated in Figure \ref{fig.Neps} to be $0.91$.

%
%

\subsection{Effect of the observable}
\begin{figure}[htbp]
\centering
\includegraphics[width=0.75\textwidth]{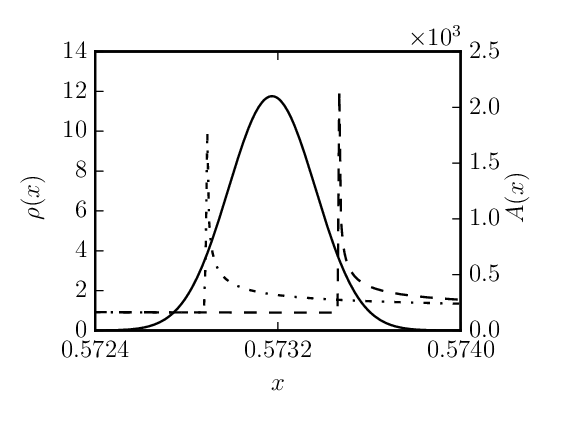}
\caption{Gaussian observable probing the displacement of the $n=11$th spike upon perturbing the logistic map (\ref{e.log}) from $a=3.8(1-\eps)$ to $a=3.8(1 +\eps)$ with $\eps=2.5 \times 10^{-6}$. The dash-dotted line shows the invariant measure at $a=3.8(1-\eps)$ and the dashed line shows the invariant measure at $a=3.8(1+\eps)$.}
\label{fig.Gauss}
\end{figure}

In the previous sections we presented results for a global observable $A(x)=x$ which probes the dynamics over the full support of the invariant measure. The breakdown is caused by the spikes in the a.c.i.m. and their rapid displacement under perturbation. It is therefore natural to expect that observables which locally probe the displacement require less data to see the breakdown of linear response. We now consider localized observables
\begin{align*}
A(x)= \exp\left(-\frac{(x-x_s)^2}{2 w^2}\right)\; .
\end{align*}
This observable allows us to probe the local non-smooth behavior of a spike $\eta_n$ at location $x_n$ with width $w_n$. Recall the displacement length $\ell_\eps=v_n\eps$ of the spike upon perturbation with $\eps$ where $v_n\sim \alpha^n$ with $\alpha$ being the average expansion rate. Hence, for $x_s\approx x_n$ and $w_n \approx w \le w^\star:=\ell_\eps/2$ the spike $\eta_n$ can be resolved by the observable and the displacement will be detectable when it leaves the effective support of the observable upon perturbation by $\eps$. An example of such a judiciously chosen Gaussian observable is given in Figure~\ref{fig.Gauss}. 

\begin{figure}[htbp]
\centering
\includegraphics[width=0.7\linewidth]{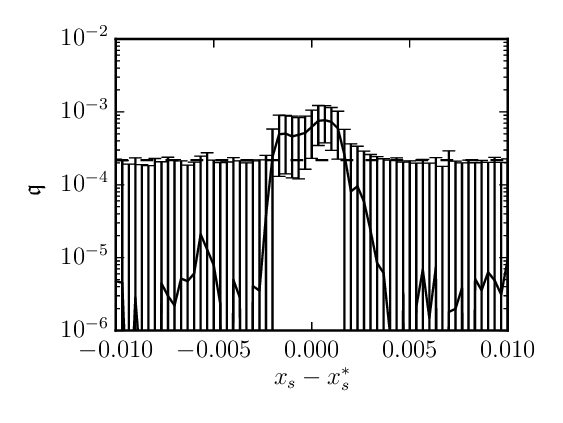}\\
\includegraphics[width=0.7\linewidth]{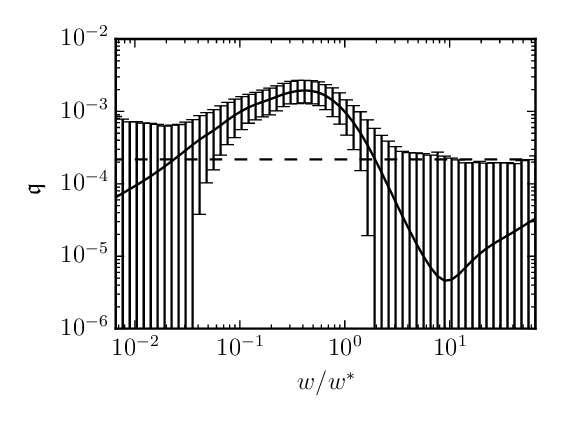}
\caption{Breakdown parameter $\hat{\q}$ (solid line) for a Gaussian observable, for the logistic map with $a=3.8(1+\varepsilon_i)$
with $d\eps=10^{-6}$ and $N=5\times10^5$. Top: as a function of the location of the center $x_s$ of the observable with $x^\star_s:=x_{11}$ the location of the $n=11$th spike. Here $w=w^\star$ is fixed to equal half the displacement length of the $11$th spike. Bottom: as a function of the width ratio $w/w^\star$ with $x_s = x_s^\star$ fixed.\\ The error bars show the two-sided $90\%$ prediction interval for $\q$ as estimated from $K=200$ realizations differing in the initial conditions. The dashed line shows $\q_\alpha$ for $\alpha=0.05$.}
\label{fig.Agauss1}
\end{figure}

The effect of a localized observable on the ability to detect breakdown of linear response is illustrated in Figure~\ref{fig.Agauss1}. We performed two sets of simulations. In the first we fixed the characteristic scale of the observable $w=w^\star$ to equal half the displacement length of the $n=11$th spike, $w^\star\approx 1.9\times 10^{-4}$, 
and varied the centre $x_s$ of the observable. A clear peak of statistically significant values of the breakdown parameter $\hat\q$ above a significance level with $p=0.05$ is obtained for $x_s$ close to the location of the $11$th spike at $x_s^\star:=x_{11}=0.573$. 
Note that the size of the displacement window within which breakdown is detectable corresponds roughly to the displacement width of the spike (cf. Figure~\ref{fig.Gauss}). In a second set of simulations we centered the Gaussian observable at the location of the $n=11$th spike with $x_s=x_s^\star$ and varied the observational scale $w$. Again, a pronounced peak of the expected value of the breakdown parameter $\hat{\q}$ is seen above the significance level for $w=w^\star$. The maximum is not obtained exactly at the estimated value of $w^\star$ due to approximations made when relating $v_n$ for finite $n$ to its asymptotic value $\alpha^n$.

\begin{figure}[htbp]
\centering
\includegraphics[width=0.7\linewidth]{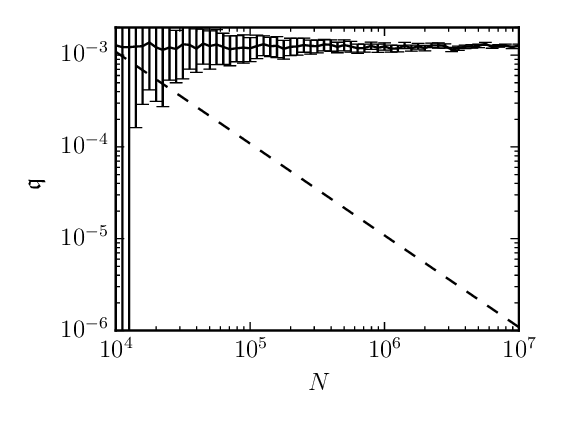}
\caption{Breakdown parameter $\hat{\q}$ (solid line) as a function of the data size $N$ for the logistic map with fixed range of perturbations $a=3.8(1+\varepsilon)$ with 
$d\eps=10^{-6}$ and Gaussian observable with width $w=w^\star$ and location $x_s=x_s^\star$ tuned to capture the displacement of the $n=11$th spike.\\ The error bars show the two-sided $90\%$ prediction interval for $\q$ as estimated from $K=200$ realizations differing in the initial conditions. The dashed line shows $\q_\alpha$ for $\alpha=0.05$.}
\label{fig.qNGauss}
\end{figure}

\begin{figure}[htbp]
\centering
\includegraphics[width=0.7\linewidth]{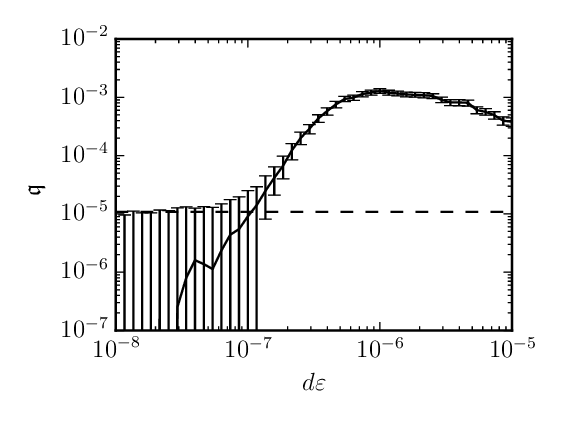}
\caption{Breakdown parameter $\hat\q$ (solid line) as a function of the perturbation size $d\eps$ for fixed $N=10^6$ 
for the logistic map (\ref{e.log}) with $a=3.8(1+\varepsilon)$ for a Gaussian observable with width $w=w^\star$ and location $x_s=x_s^\star$ tuned to capture the displacement of the $n=11$th spike.\\ The error bars show the two-sided $90\%$ prediction interval for $\q$ as estimated from $K=200$ realizations differing in the initial conditions. The dashed line shows $\q_\alpha$ for $\alpha=0.05$.}
\label{fig.qepsGauss}
\end{figure}

Figure~\ref{fig.qNGauss} shows the breakdown parameter $\hat{\q}$ as a function of the available data length $N$ for a given perturbation size $d\eps=10^{-6}$ for a Gaussian observable where $x_s=x_s^\star$ and  $w=w^\star$ is chosen to focus on the displacement of the $11$th spike. It is revealed that a time series with only $N=30,000$ is needed to reliably detect breakdown of linear response; this should be compared to the required length of $N=600,000$ when an observable $A(x)=x$ is used (cf.  Figure~\ref{fig.qN}). Similarly, Figure~\ref{fig.qepsGauss} shows the breakdown parameter $\hat\q$ as a function of the perturbation size (here the perturbation interval $d \eps$) for fixed data length $N=10^6$ for a Gaussian observable with finely tuned $w=w^\star$ and $x_s=x_s^\star$. Breakdown is reliably detected for perturbation sizes with $d\eps \ge 1.2\times 10^{-7}$; for an observable $A(x)=x$ with $N=10^6$ one needs larger perturbation sizes with $d\eps\ge 8\times 10^{-7}$ to detect breakdown (cf. Figure~\ref{fig.qeps}).

\begin{figure}[htbp]
\centering
\includegraphics[width=0.7\linewidth]{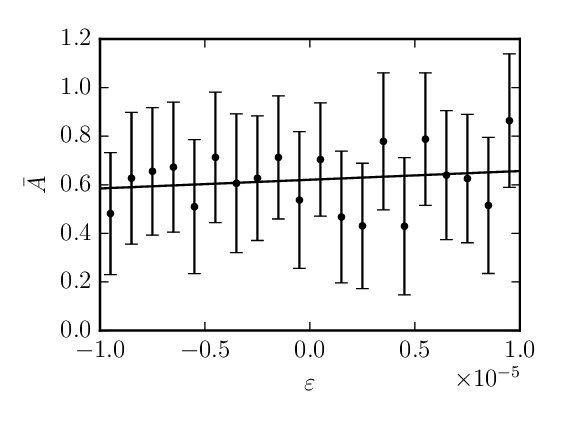}\\
\includegraphics[width=0.7\linewidth]{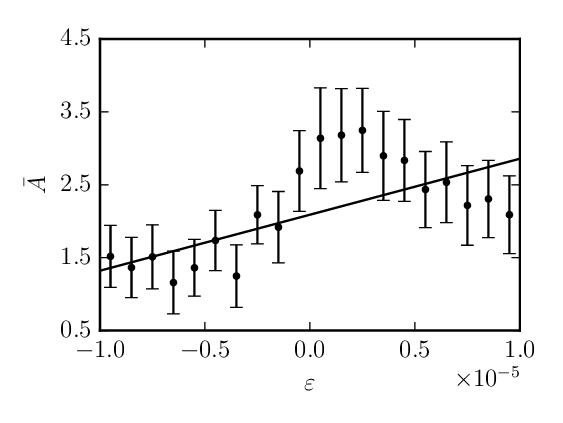}
\caption{The observed sample average $\bar{A}$ as a function of the perturbation size $\eps$ for an observable $A(x) = \exp((x-x_s)^2/(2 w^2))$ with $N=5\times 10^4$ where $w=w^\star$ is half the displacement length of the $n=11$th spike, and a linear fit (solid line). Top: when the location of the observable $x_s$ is not centered at the location of the spike $x_s^\star$. Bottom: when the location of the observable $x_s$ is centered at the location of the spike $x_s^\star$. Here $\bar A$ was obtained from a single simulation.}
\label{fig.Agauss2}
\end{figure}

The preceding discussion indicates that detailed knowledge of the underlying dynamical system (both the location of a spike $x_s^\star$ and its displacement scale $w^\star$) is required for the successful detection of the breakdown of linear response given a time series of finite length $N$. In particular, these finely tuned observables depend on the magnitude of the perturbation $\eps$. Figure~\ref{fig.Agauss2} illustrates how a lack of this knowledge may indeed mislead us into deducing the validity of linear response. We show the observed sample average $\bar{A}$~\eqref{e.Aavgs} for a Gaussian observable as a function of the perturbation size $\eps$ in the case when the characteristic observational scale $w$ and the location $x_s$ are judiciously chosen to probe for a particular spike and in the case when they are not chosen to align with a spike and its displacement length. In the latter case the existence of linear response is consistent with the observations and a scientist might be misled in believing in a linear relationship between the perturbation and the response. If, however, the location and scale of the observable are tuned to match a particular spike and its least rapid displacement of the perturbation sizes under consideration, the breakdown is clearly detectable. This, of course, as we have seen above, requires the length of the time series to be sufficiently large. The saturation of the response for sufficiently large perturbations in the case of a finely tuned localized observable (cf. bottom plot in Figure \ref{fig.Agauss2}) is an indication that the length of the time series is insufficient to detect the contribution of the other spikes to the non-smoothness of the invariant measure.

\begin{figure}[htbp]
\centering
\includegraphics[width=0.7\linewidth]{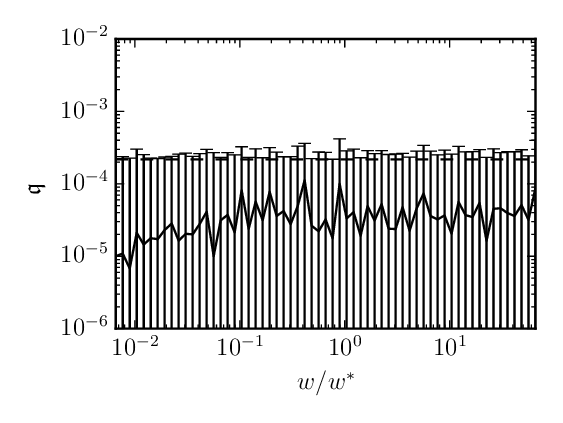}
\caption{Breakdown parameter $\hat{\q}$ (solid line) for the logistic map with $a=3.8(1+\varepsilon)$ with $d\eps=10^{-6}$ as a function of the wavelength $w$ of an observable $A(x)=\cos(2\pi(x-x_s^\star)/w)$ where $x_s^\star$ and $w^\star$ denote the location and half the displacement length of the $n=11$th spike.\\ The error bars show the two-sided $90\%$ prediction interval for $\q$ as estimated from $K=200$ realizations differing in the initial conditions. The dashed line shows $\q_\alpha$ for $\alpha=0.05$.}
\label{fig.Asine}
\end{figure}

It is pertinent to state that the mere inclusion of a scale $w=w^\star$ in the observable to probe the non-smooth dynamic behavior of the spikes of the invariant measure is not sufficient to detect breakdown for smaller values of $N$ (for fixed $\eps$) or for smaller perturbation sizes $\eps$ (for fixed $N$). For example, an observable $A(x)=\cos(2\pi(x-x_s)/w)$ with wave length $w=w^\star$ and $x_s=x^\star$ finely tuned to capture the displacement of the $11$th spike does not exhibit any variation of the expected value of the breakdown parameter $\hat\q$ as a function of the scale parameter $w$. 
Figure~\ref{fig.Asine} reveals that there is no peak in the breakdown parameter near $w=w^\star$ for finite $N=5\times 10^4$. The failure of the ${\rm{cosine}}$-function to enhance the detectability of breakdown of linear response, despite its wave length matching the characteristic displacement length of a particular spike, is due to the global character of the ${\rm{cosine}}$-function. Although the non-smooth behavior of the fast and narrow spike is resolved by the observable, this is swamped by the dominant contribution of the observable stemming from other parts of the a.c.i.m., in particular from the smooth background and from the larger, slower spikes.


%


\section{The fluctuation-dissipation theorem}
\label{sec-fdt}

The methods presented in Section \ref{sec-finite} are based on performing the perturbation experiment by brute force, i.e. by running a numerical experiment for a range of values of $\eps$. However, one of the aspects of linear response theory that has  attracted a lot of attention from practitioners is the fact that for many systems, if linear response holds, formulae exist (see Eqs. (\ref{e.fdt0}) and (\ref{e.fdt})) that express  the linear response in terms of properties of the unperturbed dynamical system, providing the tantalizing prospect of predicting the linear response without having to perform the kind of brute perturbation experiment used in Section \ref{sec-finite}.

In that vein, the fluctuation-dissipation theorem has been applied to various atmospheric and climate models. It has been mostly applied in the form of the so-called quasi-Gaussian approximation, where the invariant measure is assumed to be Gaussian, resulting in a response in the form of an integrated auto-covariance function \cite{Leith75}. This autocovariance function can be estimated from unperturbed model integrations or from measurements. The assumption of Gaussianity may be reasonable for some large-scale climatic observables, but it is not valid universally, for example for observables related to bi-stable subsystems such as the Kuroshio Extension or the El Ni\~no Southern Oscillation. A more general approach was taken in \cite{CooperHaynes11}, where the invariant measure was not assumed to be Gaussian, but was obtained by smoothing the observed empirical density with a smoothing kernel. Since we are dealing here with highly non-Gaussian densities, we will investigate this approach rather than imposing Gaussianity. 

In this section we consider the situation where one is unaware of the existence or absence of linear response for the system of interest, but only has access to a data set of observations of the unperturbed system. In such a case a practitioner might be led to estimate the right hand side of \eqref{e.fdt} from data and hope that the obtained quantity gives an indication of the response over a certain range of $\eps$. When linear response holds this will be the case, however here we investigate whether such an approximation of the response is possible when the response is non-differentiable. 

\begin{figure}[htbp]
\centering
\includegraphics[width=0.7\linewidth]{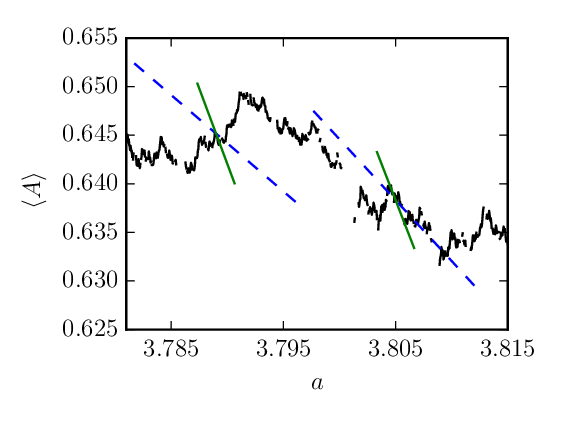}
\caption{Expectation value $\langle A\rangle$ as a function of the parameter $a$ of the logistic map (\ref{e.log}) for an observable $A(x) = x$. Gaps in the curve correspond to periodic windows. At two values of $a$ the predictions of the fluctuation dissipation theorem \eqref{e.fdt} are shown for two different kernel-smoothing widths $\omega_s$. The continuous lines (online green) are for $\omega_s = 0.005$ and the dashed lines (online blue) are for $\omega_s = 0.015$.}
\label{fig.nondiffFDT}
\end{figure}

To this end we perform a perturbation experiment for the logistic map and compare the actual response to the prediction obtained through \eqref{e.fdt}. The actual response is obtained through spectral methods \cite{Ding93,Boyd01,Trefethen13}, in order to avoid finite sample size effects. We have followed the non-parametric method based on kernel smoothing for estimating the linear response from the FDT as described in \cite{CooperHaynes11}. The density of the a.c.i.m. $\rho_0$ is smoothed by convolution with a Gaussian with smoothing width $\omega_s$. This removes the non-differentiable character of the spikes and allows the derivative to be taken in Eq. \eqref{e.fdt}.

The results of such an estimation of the linear response using the fluctuation-dissipation theorem are shown in Figure~\ref{fig.nondiffFDT}. The experiment was performed for an observable $A(x)=x$ at two different reference states $a=3.789$ and $a=3.805$ and for two different smoothing widths $\omega_s=0.015$ and $\omega_s=0.005$. It is evident that the results are sensitive to both these parameters and that the actual response is not well approximated by the slope as constructed through the FDT. 

\begin{figure}[htbp]
\centering
\includegraphics[width=0.7\linewidth]{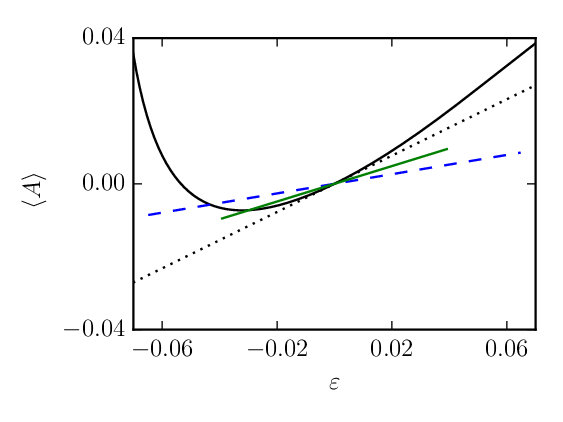}\\
\includegraphics[width=0.7\linewidth]{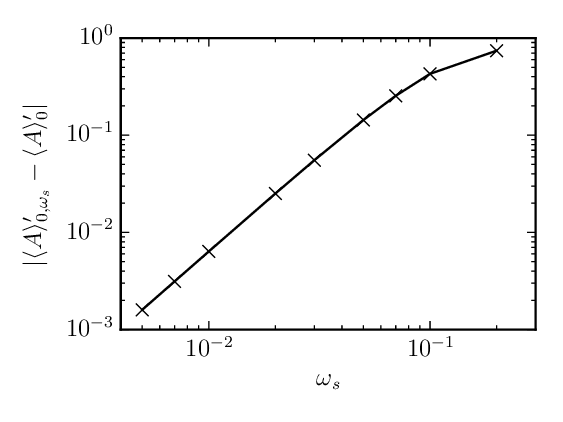}
\caption{
Top: Expectation value  $\langle A\rangle$ as a function of the perturbed conjugation of the doubling map for an observable $A(x) = \cos 2\pi x$, and perturbation $X(x) = \sin 2\pi x$. Around $\varepsilon = 0$, the predictions of the fluctuation dissipation theorem \eqref{e.fdt} are shown for two different kernel smoothing widths $\omega_s$. The continuous line (online green) is for $\omega_s = 0.05$ and the dashed line (online blue) is for $\omega_s = 0.07$. The dotted straight line shows the true linear response. Bottom: Discrepancy between the linear response estimate from kernel smoothing and the true linear response, as a function of the kernel smoothing width $\omega_s$.}
\label{fig.dblsmthFDT}
\end{figure}

We also present results showing that kernel-smoothing allows for a reliable, convergent estimation of the fluctuation-dissipation theorem in the case of a topological conjugate of the doubling map in Figure~\ref{fig.dblsmthFDT}. We use the smooth conjugation $h(x)=\frac{1}{2\pi}\cot^{-1} (0.25 + \cot 2\pi x)$, which transforms the doubling map's physical Lebesgue measure into $d\rho(x)= h'(x)dx$. We use a perturbation $X(x) = \sin 2\pi x$, and an observable $A(x) = \cos 2\pi x$. It is seen that using kernel smoothing in the fluctuation-dissipation formula (\ref{e.fdt}) approximates the true linear response well for a variety of kernel widths $\omega_s$. Furthermore the linear response estimated using kernel smoothing converges to the true linear response as $\mathcal{O} (\omega_s^2)$ as $\omega_s\to 0$. In the appendix we show analytically that the linear response as estimated using kernel smoothing converges for uniformly expanding maps to the true linear response upon decreasing smoothing width and that the error decreases with $\omega_s^2$.

\section{Summary and conclusion}
\label{sec-summary}
We have provided a detailed numerical investigation of the manifestation of breakdown of linear response caused by the non-smooth change of the invariant measure with respect to a small perturbation in a finite time series analysis. We have presented results for the logistic map for which the breakdown of linear response is analytically well understood \cite{BaladiSmania08,BaladiSmania10,Baladi14,BaladiEtAl15}.\\

The main messages which can be deduced from our results are that in order to detect the breakdown of linear response in time series of finite length, the data length needs to be sufficiently long and furthermore that the detectability of linear response strongly depends on the observable and on the perturbation size. We summarize our key findings:
\begin{itemize}
\item The amount of data $N$ required to detect a breakdown of linear response for a given perturbation size can be very large. For the logistic map with a given perturbation size of the order of $\eps={\cal{O}}(10^{-6})$ one needs at least $N=600,000$ for a smooth observable $A(x)=x$. Hence, an apparent linear response seen in a given time series might be spuriously caused by an insufficient quantity of data.
\item The smaller the perturbation size the longer the data need to be to detect a breakdown in general.
\item The global character of an observable may inhibit the detection of breakdown of linear response. For a given finite data length and given perturbation size, suitably localized observables may be needed to probe linear response. This, however, requires either detailed knowledge of the underlying dynamical system or computationally involved scans of the parameters of the observable such as its scale and its location.
\item Predicting response using the fluctuation-dissipation theorem is highly sensitive to the applied smoothing needed to assure differentiability of the density, and to the point in parameter space where the response is calculated, negating its predictive value. In the case when the FDT is valid, however, our results suggest that kernel smoothing as applied by climate scientists yields a valid approximation to the true linear response.
\end{itemize}

These findings can be taken as a word of caution for practitioners interpreting observational or numerical time series. Our results aim to narrow the gap between the body of rigorous theoretical work and the applied research done, for example, in climate science and in atmosphere and ocean dynamics.

Since we currently have no means of deciding whether the coupled atmosphere-ocean system or the whole climate system satisfies linear response theory or whether it does not, our work does not {\em{per se}} question the validity of the many results obtained using linear response theory and FDT.\\ The {\em{chaotic hypothesis}} of Gallavotti-Cohen \cite{GallavottiCohen95a,GallavottiCohen95b} is often invoked to argue that a high-dimensional chaotic physical system can be treated for all practical purposes as if it were Axiom A. It is, however, pertinent to mention, that the chaotic hypothesis only makes a statement about the existence of time averages computed with a probability distribution capturing the statistics of macroscopic observables and satisfying a large deviation law at {\em{one}} parameter value; it does not make any statement about the smoothness of the underlying probability density with respect to changes in this parameter and about whether the invariant measures of the approximating Axiom A systems at nearby parameter values are approximately linearly related, which is what is required for linear response theory. We adhere, however, to the current general belief that large complex systems with multi-scale dynamics behave as stochastic systems and therefore linear response theory is valid for large-scale observables (provided the dynamics is not close to a critical point). 

For scientists analyzing time series, we propose the following as a practical guide, which could be drawn from our work. In the case when the time series is obtained by costly numerical simulations, prohibiting the usage of very large time series, or by a limited amount of observational data, scientists could perform an ensemble of (parallel) simulations for several moderate data lengths $N$ or of subsamples. If the number of realizations which produce values of the sample mean of the breakdown parameter $\hat\q$ exceeding the corresponding threshold value for a specified significance level increases with increasing data length $N$, then this indicates breakdown of linear response as for example seen in Figure~\ref{fig.qN}. The figure suggests that another indication for a finite value of $\E \q$ and breakdown of linear response is the case where $\hat\q$ either increases or saturates over the available range of $N$. These two criteria, although far from being decisive, may be used as sufficient conditions for breakdown of linear response.



\section*{Appendix}
We prove the convergence of the estimated linear response using kernel smoothing to the true linear response in the case of a uniformly expanding one-dimensional map on a compact manifold $\Lambda$, under the assumptions that the invariant measure $\rho$ is $C^4$, and the smoothing kernel has a zero first moment. We further assume that our observable $A(x)$ is an $L^1$ function. For simplicity we assume the map has two branches, and each individual branch is at least $C^2$. This includes the doubling map as discussed in the main part.

We recall the fluctuation-dissipation theorem (\ref{e.fdt})
\begin{align}
\langle A\rangle'_\ve = -\sum_{n=0}^{\infty}\left\langle \left(\frac{ (\rho_\ve X)'}{\rho_\ve } \right) \left(A\circ f^n_\ve - \langle A \rangle_\ve\right) \right\rangle_\ve .
\label{e.fdt2}
\end{align}
which gives the linear response in terms of correlations.

Analogously, in the case where the density is kernel smoothed, the linear response is written as
\begin{align}
\langle A\rangle'_{\ve,\omega_s} = -\sum_{n=0}^{\infty}\left\langle \left(\frac{ (\mathcal{S}\rho_\ve X)'}{\mathcal{S}\rho_\ve } \right)\left(A\circ f^n_\ve - \langle A \rangle_\ve\right) \right\rangle_\ve\label{e.fdt3},
\end{align}
where $\mathcal{S}$ is a convolution by a kernel density $\phi$ with zero first moment and variance $\omega_s^2$. In particular, we have for the kernel smoothed density
\[ 
\mathcal{S}\rho(x) = \int \rho(x-y)\phi(y) dy.
\]

Dropping the $\ve$ subscripts, the difference between the kernel smoothed and the true linear response is given by
\begin{align*}
\langle A\rangle'_{\omega_s} - \langle A\rangle' &= -\sum_{n=0}^{\infty}\left\langle \left(\frac{ (\mathcal{S}\rho X)'}{\mathcal{S}\rho } - \frac{ (\rho X)'}{\rho } \right) \left(A\circ f^n - \langle A \rangle\right) \right\rangle\\
&= -\sum_{n=0}^{\infty}\left\langle X \left(\frac{ \mathcal{S}\rho'}{\mathcal{S}\rho } - \frac{ \rho'}{\rho } \right) \left(A\circ f^n - \langle A \rangle\right) \right\rangle,\\
\end{align*}
which can be bounded for $C^2$ maps with two branches by
\begin{align}
\left| \langle A\rangle'_{\omega_s} - \langle A\rangle' \right| \leq \left\| X \left(\frac{ \mathcal{S}\rho'}{\mathcal{S}\rho } - \frac{ \rho'}{\rho } \right)\right\|_{\Lip} \frac{C \|A\|_1}{1-\gamma}, \label{e.td}
\end{align}
for some $C >0$, $\gamma\in (0,1)$ independent of $A$ \cite{melbournelectures}. Here the Lipschitz-norm satisfies 
\[ \| B \|_{\Lip} = \Lip B + \| B \|_\infty \leq \| B' \|_\infty + \| B\|_\infty. \]

The right-hand-side of (\ref{e.td}) can be further bounded by using
\begin{align} 
\left\| X \left(\frac{ \mathcal{S}\rho'}{\mathcal{S}\rho } - \frac{ \rho'}{\rho } \right)\right\|_{\Lip} &\leq (\| X' \|_\infty + \| X \|_\infty) \left\|\frac{ \mathcal{S}\rho'}{\mathcal{S}\rho } - \frac{ \rho'}{\rho } \right\|_\infty \nonumber\\
	& \quad + \| X \|_\infty \left\|\left(\frac{ \mathcal{S}\rho'}{\mathcal{S}\rho } - \frac{ \rho'}{\rho } \right)'\right\|_\infty\; .
\label{e.tsd}
\end{align}
Furthermore, we can bound
\begin{align} \left\|\frac{ \mathcal{S}\rho'}{\mathcal{S}\rho } - \frac{ \rho'}{\rho } \right\|_\infty \leq \frac{\| \rho \mathcal{S}\rho' - \rho' \mathcal{S}\rho\|_\infty}{\inf\rho \inf\mathcal{S}\rho } \label{e.sd0} \end{align}
and
\newcommand{\inm}[1]{\left\| #1 \right\|_\infty}
\newcommand{\mS}{\mathcal{S}}
\begin{align} \left\|\left(\frac{ \mS\rho'}{\mS\rho } - \frac{ \rho'}{\rho } \right)'\right\|_\infty &\leq \frac{\| \rho \mS\rho' - \rho' \mS\rho\|_\infty (\inm{\rho}\inm{\mS \rho'}  + \inm{\mS \rho}\inm{\rho'})}{(\inf\rho \inf \mS\rho )^2} \nonumber\\
&\quad+ \frac{\inm{\rho \mS\rho'' - \rho'' \mS\rho}}{\inf\rho \inf\mathcal{S}\rho}\; . 
\label{e.sd1} 
\end{align}
Note that since we assume the dynamics to be uniformly expanding on a compact manifold the invariant measure is bounded away from zero with $\rho(x)>c$ for some $c>0$. Since $\phi\geq 0$, we have also ${\mathcal{S}}\rho(x) > c$, and hence, $\inf\rho >0$ and $\inf\mathcal{S}\rho>0$. 

Now for any twice-differentiable functions $p$ and $q$, we have
\[ 
(p\ \mathcal{S}q - q\ \mathcal{S}p)(x) = \int \left(p(x) q(x-y) - q(x) p(x-y) \right) \phi(y) dy.
\]

Taylor expanding $p$ and $q$ in $y$, we find 
\begin{align*}
(p\ \mathcal{S}q - q\ \mathcal{S}p)(x) 
&= 
\int \left( p(x) \left(q(x) - y q'(x) + \frac{y^2}{2} q''(\xi_1(y)) \right) \right. \\
&  \quad -
\left. q(x) \left(p(x) - y p'(x) + \frac{y^2}{2} p''(\xi_2(y)) \right) \right) \phi(y) dy\\
&= \int \left(p(x) q''(\xi_1(y)) - q(x) p''(\xi_2(y)) \right)  \frac{y^2}{2} \phi(y) dy	 \, ,
\end{align*}
where $\xi_{1,2}\in \Lambda$ and where we have used that $\phi(y)$ has a vanishing first moment.
Hence we can bound
\[ 
\|p\ \mathcal{S}q - q\ \mathcal{S}p\|_\infty \leq \frac{1}{2} \left(\inm{p} \inm{q''} + \inm{p''}\inm{q} \right) \omega_s^2.
\]

Applying the last inequality to (\ref{e.sd0}) and (\ref{e.sd1}), we can bound the right-hand-side of (\ref{e.td}) and arrive at our final estimate for the difference between the true linear response and the kernel smoothed linear response
\[ 
\left| \langle A\rangle'_{\ve,\omega_s} - \langle A\rangle'_\ve \right| < Q\|A\|_1 \, \omega_s^2 , 
\]
for some $Q > 0$ independent of $\omega_s$ and $A$. Hence the difference between the true linear response and the kernel smoothed linear response scales with the square of the kernel width as observed in Figure~\ref{fig.dblsmthFDT}. We remark that this proof can be readily extended to the case where the kernel depends on $x$.

\section*{Acknowledgement}
We would like to thank Viviane Baladi for interesting discussions. J. Wouters' research was funded by the European Community's Seventh Framework Programme (FP7/2007-2013) under grant agreement n\ensuremath{{}^{\circ}} PIOF-GA-2013-626210.

\section*{References}
\bibliographystyle{natbib}


\end{document}